## CONDENSED-MATTER SPECTROSCOPY

# Emission and Excitation Spectra of ZnO:Ga and ZnO:Ga,N Ceramics

P. A. Rodnyĭ[a], I. V. Khodyuk[a], E. I. Gorokhova[b], S. B. Mikhrin[a], and P. Dorenbos[c]

[a] *St. Petersburg State Technical University, St. Petersburg, 195251 Russia*
[b] *All-Russia Research Center Vavilov State Optical Institute, St. Petersburg, 192171 Russia*
[c] *Faculty of Applied Sciences, Delft University of Technology, 2629 JB Delft, The Netherlands*
e-mail: Rodnyi@tuexph.stu.neva.ru
Received June 17, 2008

**Abstract**—The spectral characteristics of ZnO:Ga and ZnO:Ga,N ceramics prepared by uniaxial hot pressing have been investigated. At room temperature, the edge (exciton) band at 3.12 eV dominates in the luminescence spectra of ZnO:Ga, while a wide luminescence band at 2.37 eV, which is likely to be due to zinc vacancies, is observed in the spectra of ZnO:Ga,N. Upon heating, the edge band maximum shifts to lower energies and the bandwidth increases. The extrapolated position of the edge-band maximum at zero temperature, $E_m(0) = 3.367 \pm 0.005$ eV, is in agreement with the data for thin zinc oxide films. The luminescence excitation spectra in the range from 3 to 6.5 eV are reported and the mechanism of energy transfer to excitons and luminescence centers is considered.

## INTRODUCTION

Zinc oxide is of great interest for researchers due to its unique properties. When the degree of ionic bonding is significant, ZnO is a semiconductor material, in which Wannier–Mott excitons manifest themselves as a narrow emission band near the fundamental absorption edge. Concerning the conductivity, ZnO can be transformed from an insulator (in the inactivated state) to a typical semiconductor by introducing IIIa-group impurity elements: Al, Ga, and In [1]. Prospects of applying zinc oxide in optical contacts, solar cells, and spintronics are considered. The characteristics of zinc oxide have been most completely described in reviews [1, 2]. Under standard conditions, ZnO has a hexagonal wurtzite structure, in which each $O^{2-}$ ion is coordinated by a tetrahedron composed of four $Zn^{2+}$ ions. The lattice constants are $a = 3.2497$ Å and $c = 5.2069$ Å; their ratio $c/a = 1.602$ is close to ideal. In view of the large degree of ionic bonding, 72-meV optical phonons play a key role in ZnO. Two emission bands are generally detected in all diverse forms of ZnO (single crystals, thin films, whiskers, nanocrystals, needles, etc.), i.e., a short-wavelength band near the absorption edge (the luminescence edge) and a long-wavelength (green) band, which will be referred to as the intraband luminescence. The edge luminescence has the exciton nature, while the intraband luminescence is due to the presence of oxygen or zinc vacancies [1] or residual impurities [3]. The edge deexcitation time lies in the nanosecond or subnanosecond range; therefore, it is most important for high-speed devices (lasers, scintillators, phosphors).

Gallium-doped zinc oxide is used in the form of thin-film coatings in scintillation detectors to detect α particles [4]. To measure X rays and γ rays, ceramics based on zinc oxide have been developed [5, 6]. It is important that ZnO has a relatively narrow (for scintillators) band gap ($E_g = 3.37$ eV) because the conversion efficiency of scintillators increases with a decrease in the band gap. In addition, the exciton binding energy in ZnO (60 meV) in higher by a factor of 2.4 than the thermal energy $kT$ (for $T \approx 290$ K), as a result of which the edge luminescence is strong at room temperature. For scintillation detectors, the following ZnO parameters are important: transparency in the visible spectral range, good thermal and mechanical properties, sufficiently high density (5.61 g/cm$^3$), and high radiation resistance.

In this paper, we report the results of studying the luminescence characteristics of the ZnO:Ga and ZnO:Ga,N ceramics obtained by uniaxial hot pressing. The optical, X-ray diffraction, and scintillation characteristics of the samples were described in [7].

## EXPERIMENTAL

The ceramics were prepared by uniaxial hot pressing [6, 7] from specially purified initial powders (Alfa Aesar production) in the form of 24-mm disks with a thickness of 1.5 mm (after polishing). A mixture of ZnO and Ga$_2$O$_3$ powders was used to prepare ZnO:Ga ceramics. In addition, we attempted to introduce nitrogen coactivator into ZnO:Ga. In contrast to gallium, which forms donor levels in ZnO, nitrogen generates shallow acceptor levels [1, 8]. It is believed that, due to the donor–acceptor recombination, the edge band in the samples activated with gallium and nitrogen should be shifted to longer wavelengths in comparison with that in ZnO:Ga. ZnO:Ga,N ceramics were prepared from a mixture of ZnO and $Ga(NO_3)_3$ powders. On the basis of the previous experiments on ceramics synthesis [6] and the data on the powder [8, 9] and single-crystal [10] ZnO:Ga, the gallium concentration in the ZnO:Ga and ZnO:Ga,N samples was chosen to be 0.075 wt %, which corresponds to ~$1.5 \times 10^{20}$ $cm^{-3}$.

The X-ray luminescence spectra were measured using an X-ray tube with a copper anode, operating at 55 kV and 40 mA. To cut off the X-ray soft component, we applied a 3-mm thick Al filter. The emission spectra were measured on a VM504 monochromator (Acton Research) with a resolution of 1 nm (diffraction grating 1200 lines/mm). An R934-04 photoelectron multiplier (Hamamatsu) was used as a photodetector. All measured spectral curves were corrected taking into account the multiplier sensitivity and the transmission capacity of the monochromator at different wavelengths.

The excitation spectra were measured in the UV spectral region; the UV source was an E7536 CW Xe lamp with a power of 150 W (Hamamatsu). The excitation scheme contained an ARC monochromator (model VM502, 1200 lines/mm (slit 0.3 mm), with an $Al/MgF_2$-coated mirror), $MgF_2$ window, and focusing elements. The detection scheme included a Macan 910 monochromator (resolution 8 nm, 1200 lines/mm) and an XP2254/B photoelectron multiplier (Philips). All recorded spectra were corrected taking into account the emitter intensities at different wavelengths.

## RESULTS

Figure 1 shows how the shape and position of the ceramics luminescence spectra depend on temperature. The integrated intensity (specific light yield) of the samples turned out to be the same; however, the emission in ZnO:Ga ceramics was concentrated predominantly in the short-wavelength region, and ZnO:Ga,N exhibits dominant intraband luminescence. In ZnO:Ga, the intraband band peak shifts from 2.05 eV at 78 K to 2.3 eV at 300 K (Fig. 1a). The wide intraband band in ZnO:Ga,N peaks at 2.37 eV; the peak position is temperature-independent (Fig. 1b). The peak positions of the edge-luminescence bands in ZnO:Ga and ZnO:Ga,N were found to be the same, i.e., 3.12 eV at 300 K, which indicates that the expected long-wavelength shift due the acceptor (nitrogen) doping was not observed. Moreover, the edge-band peak in the doped samples is shifted with respect to that for undoped ZnO ceramics [6] by only ~10 meV. The integrated edge-luminescence intensity for ZnO:Ga at 78 K is higher than at 300 K by a factor of 3; at $T > 300$ K, it exceeds the intraband luminescence intensity.

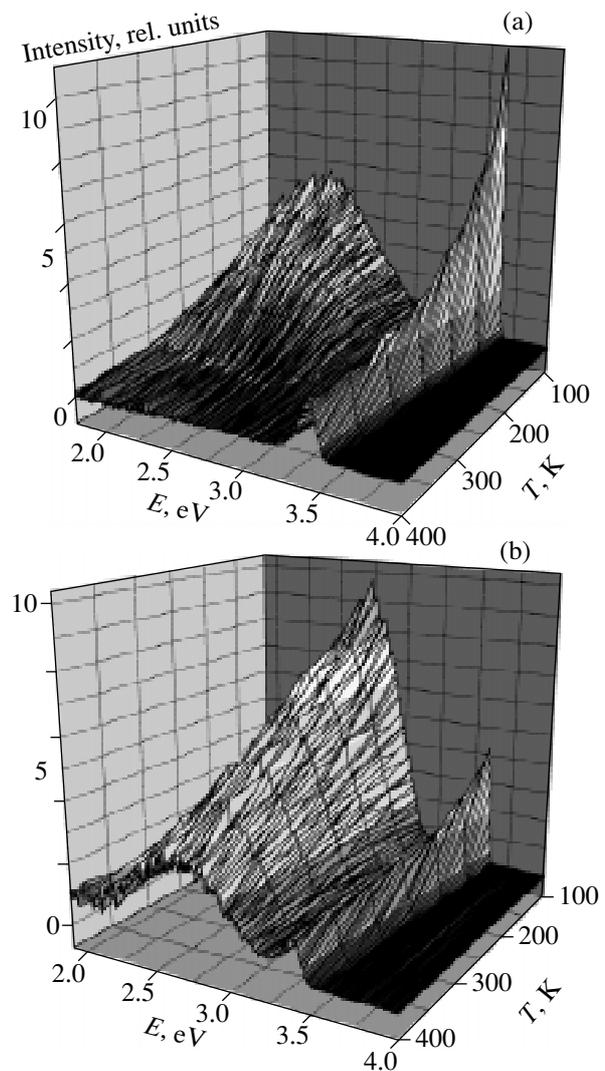

**Fig. 1.** Dependences of the luminescence intensities of (a) ZnO:Ga and (b) ZnO:Ga,N ceramics in the energy–temperature coordinates.

The edge-luminescence spectra of the ZnO:Ga and ZnO:Ga,N ceramics at different temperatures are shown in Fig. 2 (the spectra were measured at temperatures from 78 to 600 K with a step of 25 K; to avoid cumbersomeness, only part of them are shown). It can be seen that, with an increase in temperature, the peak of the edge-luminescence band shifts to lower energies (undergoes a red shift), and the bandwidth increases. The arrow in Fig. 2 indicates a feature in the spectrum at 78 K, which is due to the contribution of longitudinal optical (LO) phonons to the luminescence.

The temperature dependence of the position of the edge-luminescence peak for the ZnO:Ga ceramics is shown in Fig. 3 (a similar dependence was obtained for ZnO:Ga,N). This dependence has two specific features: a large peak shift (by 0.4 eV) in the temperature range of 100–600 K and almost linear temperature dependence of the peak position.

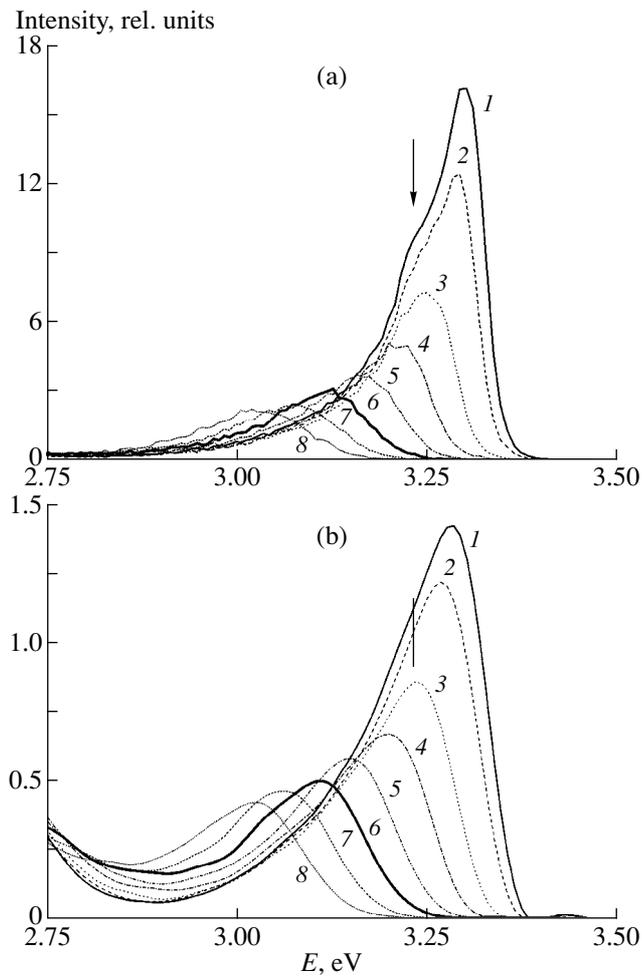

**Fig. 2.** Edge-luminescence spectra of the (a) ZnO:Ga and (b) ZnO:Ga,N ceramics at temperatures $T$ = (*1*) 78, (*2*) 100, (*3*) 150, (*4*) 200, (*5*) 250, (*6*) 300, (*7*) 350, and (*8*) 400 K.

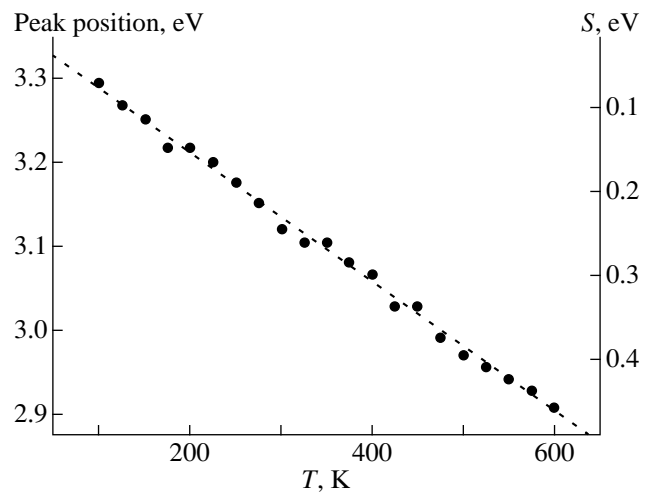

**Fig. 3.** Temperature dependence of the position of the edge-luminescence peak for the ZnO:Ga ceramics. The luminescence Stokes shift $S$ is plotted on the right ordinate axis.

The temperature dependence of the half-width at half maximum ($\Gamma$) for the edge-luminescence band of ZnO:Ga is shown in Fig. 4. The data obtained have a large spread because they were derived directly from the experimental curves (Fig. 2). These spectral curves include (i) phonon replicas, which expand the low-energy band edge, and (ii) the fundamental absorption component, which cuts off the high-energy edge of the luminescence band. In addition, the temperature decay of the luminescence intensity is imposed on these effects, as a result of which the true value is difficult to determine.

Figure 5 shows the dependences of the quantum yields of the edge and intraband luminescence on the incident photon energy $h\nu_{exc}$ for the ZnO:Ga and ZnO:Ga,N ceramics at 290 K. The intraband luminescence is effectively excited in the range of exciton generation (the band peaking at 3.17 eV) and is not excited near the interband transition, i.e., at $h\nu_{exc} > E_g = 3.37$ eV (Fig. 5, curves *1*, *2*). The luminescence edge is excited at incident photon energies slightly exceeding the zinc oxide band gap (the inset in Fig. 5 shows the dependence for the ZnO:Ga ceramics; for ZnO:Ga,N, the dependence is the same, and the curves coincide). Both the edge and intraband bands are barely excited in the range of 3.5–10 eV.

## DISCUSSION

The intraband luminescence in ZnO may be due to zinc vacancies $V_{Zn}$ [1], oxygen vacancies $V_O$ [11], antisite zinc $O_{Zn}$ [12], and other centers. A recent investigation [13] showed that the luminescence band peaking at 2.35 eV is due to $V_{Zn}$ centers, while oxygen vacancies are responsible for the shorter-wavelength radiation (2.53 eV). The position of the wide long-wavelength luminescence band (Fig. 1) in the spectra of the ceramics studied suggests that this band is due to $V_{Zn}$ centers. Obviously, in our case, the introduction of $Ga_2O_3$ into ZnO decreases the content of zinc vacancies in the sample and enhances the edge luminescence.

The red shift of the edge luminescence in ZnO is generally attributed to the decrease in the crystal band gap upon cooling. The ZnO band gap decreases by 65 meV in the temperature range from 100 to 300 K [1], the red shift for single crystals [14] and films [15] is approximately the same. For the ceramics, the corresponding value is $\Delta E_m = 0.18$ eV. The shift of the high-energy band edge in the range 100–300 K is also large, i.e., 0.11 eV (Fig. 3). Even taking into account the Urbach absorption tails, one cannot explain such a large red (Stokes) shift in the spectra of ZnO:Ga and ZnO:Ga,N ceramics. In addition, the edge-luminescence bands do not have a sharp high-energy wing at $T \geq 250$ K; i.e., they are less affected by the intrinsic absorption in ZnO. The intrinsic absorption is significant for temperatures $T < 250$ K, at which the edge-luminescence band is asymmetric (Fig. 2). At 78 K, the luminescence band of ZnO:Ga (Fig. 2b, upper curve) can be approximated by two Gaussians peaking at $E_{m1} = 3.302$ eV and $E_{m2} = 3.230$ eV. The former and latter bands should be attributed to the excitons localized on donors ($D,X$ centers) and excitons bound with optical phonons ($D,X$ –1LO) because $E_{m1} - E_{m2} = 72$ meV.

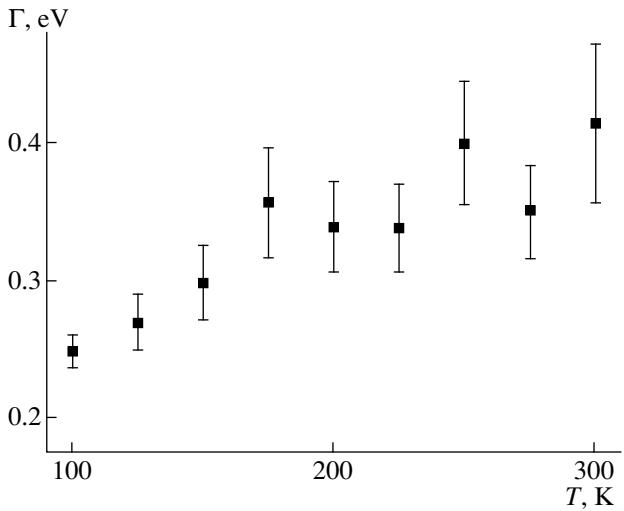

**Fig. 4.** Temperature dependence of the half-width at half-maximum, $\Gamma$, of the edge-luminescence band of ZnO:Ga. The errors are shown by vertical lines.

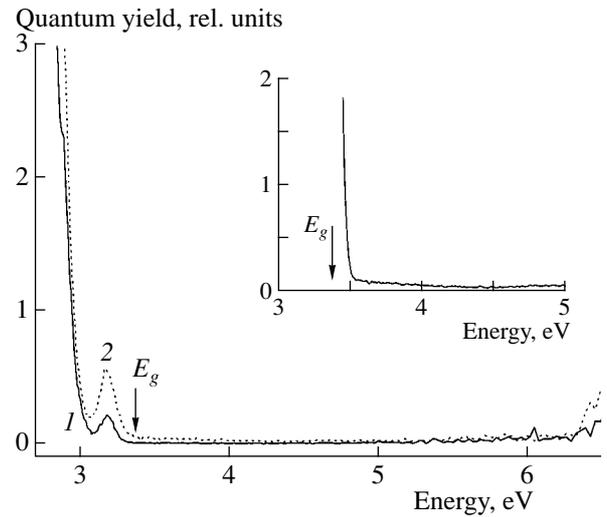

**Fig. 5.** Intraband-luminescence excitation spectra of the (*1*) ZnO:Ga,N and (*2*) ZnO:Ga (*2*) ceramics at $T = 290$ K. The inset shows the edge-luminescence excitation spectrum of ZnO:Ga.

The specific features of the temperature changes in the edge-band parameters—large Stokes shift and linear dependence of the peak position—should be attributed to the presence of several gallium levels near the top of the conduction band. It is known that gallium in low concentrations ($\sim 10^{18}$ cm$^{-3}$) forms single levels near the bottom of the conduction band in ZnO, whereas a system of levels is formed at high concentrations ($\geq 10^{20}$ cm$^{-3}$) [9]. The shift of the edge-band peak (Fig. 3) can be caused by to the successive deoccupation of gallium levels in ZnO:Ga. Upon heating, the electrons from high-lying levels are thermally excited to the conduction band, and low-lying levels are involved in emission. This mechanism is confirmed by the exponential decrease in the edge-luminescence intensity with an increase in temperature (Fig. 1). Naturally, the temperature change in the band gap and the shift of the fundamental absorption edge of ZnO contribute to the shift of the edge-band peak.

The dependence $E_m = E_m(T)$ for ZnO crystals [14] and films [15, 16] differs from that obtained by us for the ZnO:Ga ceramics (Fig. 3). In the case under consideration, the following linear dependence is observed for the red shift[1]:

$$E_m(T) = E_m(0) - \alpha T.$$

Using the dependence in Fig. 3, we obtain $E_m(0) = 3.367 \pm 0.005$ eV; $\alpha = 0.774$ meV/K. The position of the edge-band peak at zero temperature, $E_m(0)$, is consistent with the energy of the excitons localized at donors ($D, X$) in single crystals (3.360 eV) [14] and films (3.3651 eV) [15]. The coefficient $\alpha$ for ceramics is between those for $D$ and $X$ excitons in single crystals ($\alpha = 0.67$ meV/K) and nanocrystals ($\alpha = 1.14$ meV/K) [14].

The Stokes shift of the edge luminescence at room temperature is 0.24 eV (Fig. 3). For comparison, the Stokes shift in thin ZnO:Ga films with a gallium concentration of $10^{20}$ cm$^{-3}$ is ~0.2 eV at room temperature [9]. The change in the edge-band half-width at half-maximum cannot be described by the generally accepted dependence [14] due to the spread in $\Gamma$ and absence of low-temperature ($T < 78$ K) experimental data. We can only state that $\Gamma$ for ceramics changes from 0.25 to 0.4 eV in the temperature range of 100–300 K. For comparison, in ZnO:Ga films with a gallium concentration of $1.5 \times 10^{20}$ cm$^{-3}$, $\Gamma = 195$ meV at 293 K [9].

Intraband luminescence is effectively excited in the range of exciton generation (the band peaking at 3.17 eV in Fig. 5). This fact indicates that intraband luminescence is due to the generation of excitons, which then radiatively annihilate at luminescence ($V_{Zn}$) centers. Note that the conventional recombination mechanism of intraband luminescence was accepted in a number of studies; i.e., the recombination of conduction-band electrons with $V_{Zn}$ or $V_O$ centers was considered [11, 16]. The sharp increase in the luminescence quantum yield at $h\nu_{exc} < 3$ eV (Fig. 5, curves *1*, *2*) is obviously caused by direct excitation of luminescence ($V_{Zn}$) centers.

---

[1] Generally, the formula [15] $E_m(T) = E_m(0) - [\alpha T/(1 + \beta/T)]$ is used for the red shift. It is likely that the ratio $\beta/T$ does not differ much from unity in our case.

The edge-luminescence excitation spectrum (Fig. 5, inset) shows that excitons in ZnO:Ga are formed upon generation of the so-called genetic electron–hole pairs and under direct excitation of exciton states. If the energy of generated electrons exceeds ~0.15 eV (with respect to the bottom of the conduction band), they cannot be excited to gallium levels and are apparently captured by traps.

Note that an increase in the quantum yield of the edge and intraband luminescence was observed at energies $h\nu_{exc} > 10.2$ eV (the corresponding curves are not reported because the deuterium lamp used by us has a low intensity at $h\nu_{exc} > 10.5$ eV, and the data are not reliable). A similar increase in the quantum yield was observed in a ZnO:Ga single crystal [10]; it was attributed to the electron transitions from the Zn 3*d* levels to the conduction band. An increase in the edge-luminescence quantum yield at $h\nu_{exc} > 10$ eV was detected in undoped ZnO ceramics [17]. Therefore, this increase is a fundamental feature of the base (zinc oxide).

## CONCLUSIONS

At room temperature, the edge- (exciton-) luminescence band peaking at 3.12 eV dominates in ZnO:Γ. In ZnO:Ga,N, an intraband-luminescence band at 2.37 eV is detected, which is apparently due to zinc vacancies. Upon heating, the edge-band peak shifts to lower energies and the bandwidth increases. For ceramics (in contrast to crystals), a linear temperature dependence (with the slope $\alpha = 0.774$ meV/K) of the edge-band peak position was recorded. The linear dependence $E_m(T)$ is the result of superposition of several processes. The position of the edge-band peak at zero temperature, $E_m(0) = 3.367 \pm 0.005$ eV, obtained by extrapolation, is in agreement with the data for thin films and crystals.

The luminescence excitation spectra suggest that intraband luminescence is excited through exciton states. The edge luminescence arises upon direct exciton generation and formation of electron–hole pairs with energies slightly exceeding the ZnO band gap.

## REFERENCES

1. U. Orgur, Ya. I. Alivov, C. Liu, et al., J. Appl. Phys. **98**, 041301 (2005).
2. B. K. Meyer, H. Alves, D. M. Hofmann, et al., Phys. Status Solidi B **241**, 231 (2004).
3. Ya. I. Alivov, M. V. Chukichev, and V. A. Nikitenko, Fiz. Tekh. Poluprovodn. (St. Petersburg) **38** (1), 34 (2004) [Semiconductors **38**, 31 (2004)].
4. A. Beyerle, J. P. Hurley, and L. Tunnele, Nucl. Instrum. Methods Phys. Res., Sect. A **299**, 458 (1990).
5. S. Takata, T. Minami, and H. Nanto, J. Lumin. **40/41**, 794 (1988).
6. V. A. Demidenko, E. I. Gorokhova, I. V. Khodyuk, et al., Radiat. Meas. **42**, 549 (2007).
7. E. I. Gorokhova, P. A. Rodnyĭ, I. V. Khodyuk, et al., Opt. Zh. (2008) (in press).
8. E. D. Bourret-Courchesne, S. E. Derenso, and M. J. Weber, Nucl. Instrum. Methods Phys. Res., Sect. A **579**, 1 (2007).
9. T. Makino, Y. Segawa, S. Yoshida, et al., Appl. Phys. Lett. **85**, 759 (2004).
10. P. A. Rodnyĭ, G. B. Stryganyuk, and I. V. Khodyuk, Opt. Spektrosk. **104** (2), 257 (2008) [Opt. Spectrosc. **104**, 210 (2008)].
11. A. van Dijken, E. A. Meulenkamp, D. Vanmaekelbergh, and A. Meijerink, J. Lumin. **90**, 123 (2000).
12. B. Lin, Z. Fu, and Y. Jia, Appl. Phys. Lett. **79**, 943 (2001).
13. T. Moe Berseth, B. G. Svenson, A. Yu. Kuznetsov, et al., Appl. Phys. Lett. **89**, 262 112 (2006).
14. V. A. Fonoberov, K. A. Alim, A. A. Balandin, et al., Phys. Rev. B: Condens. Matter Mater. Phys. **73**, 165 317 (2006).
15. H. J. Ko, Y. F. Chen, Z. Zhu, et al., Appl. Phys. Lett. **76**, 1905 (2000).
16. B. Cao, W. Cai, and H. Zeng, Appl. Phys. Lett. **88**, 161 101 (2006).
17. L. Grigorjeva, D. Millers, J. Grabis, et al., IEEE Trans. Nucl. Sci. **55** (3) (2008).